\documentclass[12pt]{book}

\usepackage[dvips]{graphicx,color}
\usepackage{makeidx,tsukuba}
\usepackage{floatflt}
\usepackage{times}
\usepackage{subfigure}

\makeauthorindex
\makeindex

\begin{document}

\BookTitle{\itshape The 28th International Cosmic Ray Conference}
\CopyRight{\copyright 2003 by Universal Academy Press, Inc.}
\pagenumbering{arabic}

\chapter{Performance of the Pierre Auger Fluorescence Detector and Analysis of 
    Well Reconstructed Events}

\author{Stefano Argir\`o $^1$ for the Pierre Auger Collaboration $^2$\\
{\it (1) University of Torino and Istituto Nazionale di Fisica Nucleare,
via Giuria 1, 10125 Torino, Italy \\
(2) Observatorio Pierre Auger, Av. San Martin Norte 304, 5613 Malargue, Argentina\\} }

\section*{Abstract}
The Pierre Auger Observatory is designed to elucidate the origin and nature 
of Ultra High Energy Cosmic Rays using a hybrid detection technique.
A first run of data taking with a prototype version of both detectors (the so
called Engineering Array) took place in 2001-2002, allowing the Collaboration
to evaluate the performance of the two detector systems and to approach an
analysis strategy. In this contribution, after a brief description of the system,
we will report  some results on the behavior of the Fluorescence Detector (FD)
Prototype. Performance studies, such as measurements of noise, sensitivity and duty 
cycle, will be presented. We will illustrate a  preliminary analysis of
selected air showers. This analysis is performed using exclusively the
information from the FD, and includes reconstruction of the shower geometry and
of the longitudinal profile.

\section*{Introduction}

The Pierre Auger Cosmic Ray observatory will be the largest cosmic ray detector ever built. Two
sites of approximately 3000 km$^2$ , one in each hemisphere,  
will be instrumented with a surface detector and a set of fluorescence detectors.
Two  fluorescence telescope units were operated from December 2001
to  March  2002 in conjunction with 32 surface detectors, the so-called 
Engineering Array. This phase of the project was aimed at proving the 
validity of the design and probing the potential of the system. In the
following we will show an analysis of the performance of the FD during this
run and demonstrate, by investigating selected events, the ability to
reconstruct geometry and the longitudinal profile of Extensive Air Showers.

\section*{System Overview}

Figure \ref{fig:tel} shows a schematic view of a fluorescence telescope unit.
An array of 20$\times$22 hexagonal photomultiplier tubes (the \emph{camera})
is mounted on a quasi-spherical support located at the focal surface of a 
segmented mirror [1].  Each PMT overlooks a region of the sky of 1.5
deg in diameter. The telescope aperture has a diameter of 2.20 m and features an
optical filter (MUG-6) to select photons in the range 300-400 nm.
A Schmidt corrector ring allows the collection area to be doubled without increasing
the effects of optical aberrations. The
Schmidt geometry results in a 30$^\circ \times$ 30$^\circ$ field of view
for each telescope unit. The final design envisages four eyes each of six
telescopes.
\begin{floatingfigure}[l]{80 mm}
\begin{center}
  \includegraphics[scale=0.32]{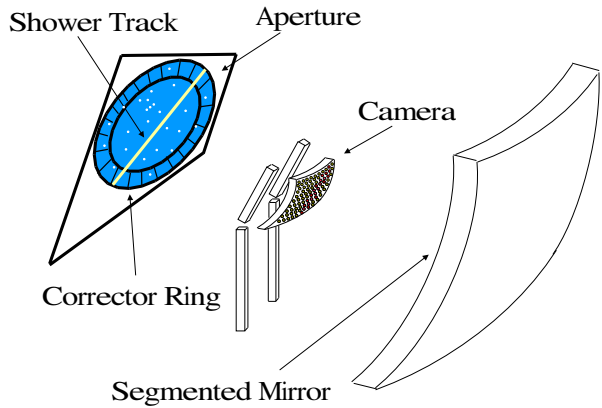}
  \caption{Schematic view of a Fluorescence Telescope unit}  \label{fig:tel}
\end{center}
\end{floatingfigure}
The current signal coming \\ from each phototube is sampled at 10 MHz with 12 bit
resolution and 15 bit dynamic range. The \emph{First Level Trigger} system 
performs a boxcar running
sum of ten samples. When the sum exceeds a threshold, the trigger is fired.
This threshold is determined by the trigger rate itself, and is regulated
to keep it close to 100 Hz. Every 
microsecond, the camera is scan\-ned for patterns of fired pixels that are 
consistent with a track induced by the fluorescence light from a sho\-wer. This is
the  \emph{Second Level Trigger}, that in the \emph{Engineering Array} run
had a rate of 0.3 Hz. It is dominated by muons hitting the
camera directly and random noise. These components, as well as lightning, are 
then filtered  by the software
\emph{Third Level Trigger}, yielding  a rate of 8$\times$10$^{-4}$ Hz (one
event every 20 mn) per telescope.

\section*{System Performance}
%
%
We will briefly report on some measured parameters that give an indication of
the performance of the optics and of the electronics. The size of the light spot on the focal  surface gives an
indication on the quality of the optics and the alignment of the mirrors. This
has been measured by placing a white screen on the camera  and then using a CCD
device to capture images from bright stars, which can effectively be considered
as point sources. Measurements taken for different star positions (therefore
different light incidence angles) show that 90\% of the light is collected
within a circle of 1.8 cm (0.65 deg) diameter.  The reflectivity of each mirror segment
was measured to be above 90\%. \\
The most important parameter of the front-end electronics  is the noise
level. The design requires a  system contribution of less than 10\% of
total  noise, which is dominated by the light background from the night sky. The
contribution  of the electronics has been measured by comparing the
fluctuation of the baseline of the sampled  signal in dark conditions,
$\sigma_{elec}$,  with that determined when the detector is exposed  to the night sky, $\sigma_{sky+elec}$.
The ratio  ${\sigma_{sky+elec}} / {\sigma_{sky}}$ , averaged over all the
pixels, gives a value of 1.06 . The noise contribution of the
sky alone is here calculated as $\sigma_{sky} = \sqrt{\sigma^2_{sky+elec} - \sigma^2_{elec}}$. 

\begin{figure}
\hspace {-2cm}
\mbox{
  \subfigure {\includegraphics[scale=0.45]{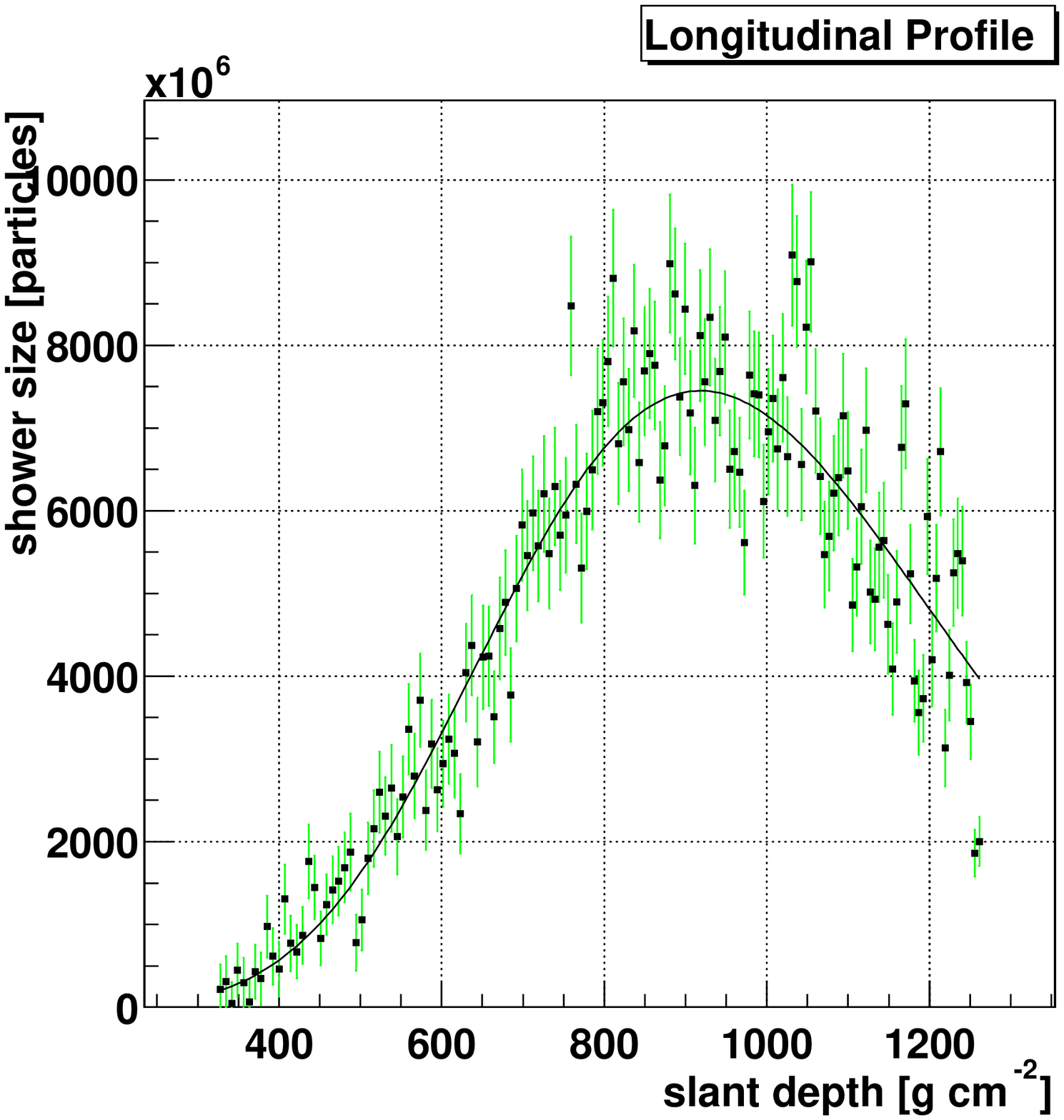}}

  \hspace {-1cm}
  \subfigure {\includegraphics[scale=0.45]{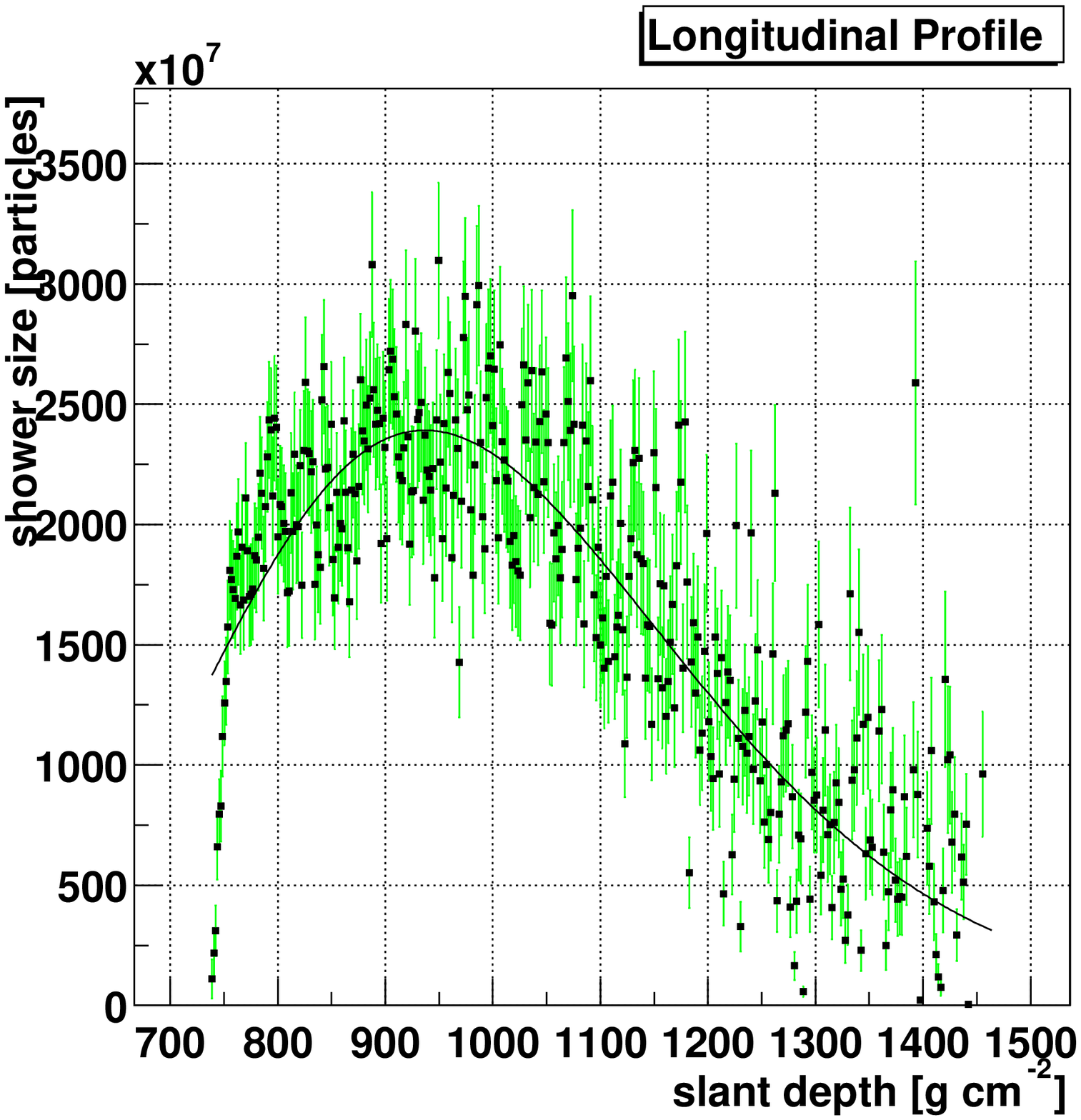}}}

\caption{\emph{Left:} Reconstructed longitudinal profile of  a shower landing
          about 13 km from the detector. The estimated energy is around 1.3$\times
	  10^{19}$ eV. The line is a fit to a Gaisser-Hillas function. 	  
	  \emph{Right:} Same for an inclined  shower landing about
	  20 km from the detector, with energy around 3.3 $\times10^{19}$ eV }
\label{fig:rec}	  

\end{figure}

The sensitivity of the system to distant showers has been estimated
using a portable laser system. This frequency-tripled YAG laser features
adjustable energy, measured by means of a radiometer applied
to a portion of the beam. The \emph{laserscope} was brought to a distance of
26 km, and the energy lowered in steps while 
the drop in FD trigger efficiency was being watched.
 Although the measurements should be extended and
improved, they  indicate a threshold around 10 EeV at 26 km.
The Fluorescence Detector prototype was run
smoothly during the period December 2001 - March 2002. During this time it has
collected over 1000 shower candidates and several hundred laser shots for detector
studies. It was operated from 30 mn  after astronomical dusk to 30 mn before
dawn, in periods when the fraction of illuminated moon was below
50\%. The duty cycle was 11\%, close to the figure foreseen for normal
operation. In the next section we will present the preliminary technique
used to reconstruct the longitudinal profile of a selected sample
of showers.

\section*{Reconstruction}
The methods to reconstruct the shower geometry with a monocular
FD alone are described in another contribution
[3], where the detector geometrical resolution is presented.
In the following we will outline the procedure used to reconstruct the
longitudinal profile and primary energy from \emph{Engineering Array} data.\\
The received light flux $S(X)$ originating in a layer between
$X$ and $X + \delta X$, where $X$ is the atmospheric depth in $g/cm^2$, 
may be approximated by:

	\begin{equation}  \label{eq:signal}
		S(X) = L(X) \frac{A}{4 \pi r^2} ~ c ~ \delta t \cdot \epsilon \cdot {\mathcal T}(r)
	\end{equation}	
where $L(X)$ is the fluorescence light isotropically emitted at the source (equal to the product
of the yield at height $h$, $F_y(h)$, and the shower size $n_e(X)$ ), $A$ is the collection area, $r$ the
observation distance,   $\delta t$ is the time the shower takes to travel from $X$ to $X+\delta X$ when
viewed from the detector ($\delta t$ depends solely on the geometry), $\epsilon$ is the collection
efficiency  and $\mathcal T(r)$ factorizes the transmission of the atmosphere. The calibration chain
[4] gives the signal in units of 370 nm equivalent photons at the aperture. Therefore, to
reconstruct the incident flux as a function of time, $S(t)$, it is sufficient to collect the calibrated signal
from all pixels within an angle $\zeta$ from the shower track. $\zeta$ is chosen so as to maximize the signal
to noise ratio. The next step is to evaluate $L(X)$ from $S(t)$. This is done by unfolding the effect of the atmosphere
transmission using, in this preliminary analysis, a standard atmosphere model to deal with Rayleigh and aerosol 
components. The fluorescence yield
$F_y(h)$ must then be estimated in order to obtain $N_e(X)$. The measured fluorescence yield from
[2] is used. The contribution of direct and scattered Cerenkov light  is then  subtracted
with an iterative procedure. The electromagnetic energy is calculated as: $E_{em} = 
\frac{E_c}{X_r}   \int n_e(X) dX$ where $n_e(X)$ is the number of electrons at depth
X[$g \cdot{cm^{-2}}$], $E_c$ is the critical energy of electrons in air and $X_r$ their radiation length.\\
We selected a sample of about 50 events recorded during the \emph{Engineering Array} run to 
exercise the presented reconstruction method. Examples are shown if Figure \ref{fig:rec} 

\section*{Conclusions}
The \emph{Engineering Array} run has proved that the Fluorescence Detector
prototype meets design specifications and performs appropriately. It was
possible to reconstruct the longitudinal profiles of several showers.

\section*{References}

\re
1.\ Auger Collaboration\ , ``Technical Design Report'', \texttt{www.auger.org}
\re
2.\ Kakimoto F. et al, 1996, Nucl. Instrum. Meth. A372, 527 \ 
\re
3.\ Privitera, P.\ , these Proceedings
\re
4.\ Roberts, M.\ , these Proceedings

\endofpaper
\end{document}